\documentclass[aps,prc,twocolumn,nofootinbib,floatfix,superscriptaddress,showpacs,amssymb,amsmath]{revtex4}

\usepackage{graphicx}

\newcommand*{\AMeV}{\textit{A}\;MeV}

\begin{document}



\title{Statistical multifragmentation features of midvelocity source
       in semiperipheral heavy-ion collisions}


%
%

\author{G. Casini},
\affiliation{Sezione INFN di Firenze, 
             Via G. Sansone 1, I-50019 Sesto Fiorentino, Italy}

\author{S. Piantelli},
\affiliation{Sezione INFN di Firenze, 
             Via G. Sansone 1, I-50019 Sesto Fiorentino, Italy}

\author{P.R. Maurenzig},
\affiliation{Sezione INFN and Universit\`a di Firenze, 
             Via G. Sansone 1, I-50019 Sesto Fiorentino, Italy}

\author{A. Olmi},
\affiliation{Sezione INFN di Firenze, 
             Via G. Sansone 1, I-50019 Sesto Fiorentino, Italy}

\author{A.S. Botvina},
\affiliation{Institute for Nuclear Research, Russian Academy of Science,
             117312 Moscow, Russia}

\date{\today}

\begin{abstract}
Some characteristics of midvelocity emissions in semiperipheral
heavy-ion collisions at Fermi energies are discussed in the framework of
a multifragmenting scenario. 
We report on binary dissipative collisions of $^{93}$Nb + $^{93}$Nb at 
38\AMeV\ in which we measured an abundant emission of particles and
fragments not originated from the usual evaporative decay of hot primary
fragments. 
We checked the compatibility of these emissions with the
multifragmentation of a source which forms in the overlap region. 
One can fairly well reproduce the  data assuming a hot and dilute
source, possibly more n-rich than the initial nuclei; the results appear
to be insensitive to the source size. 
\end{abstract}

\pacs{25.70.Lm, 25.70.Pq}

\maketitle

       \section{Introduction}

The recent literature on experimental heavy-ions physics in
(semi-)~peripheral collisions at Fermi energies 
(see e.g.~\cite{pia2,xu,mang,huda5,defil,pia6,xiao,pla}) often focuses
on the emission of particles and fragments from the phase-space region
in between the reacting nuclei, called midvelocity emission.  
This phenomenon has been already investigated by several authors, as a
function of the size of the system and of the impact parameter, but its
origin is far from being understood; in particular, it is not clear if
one deals with only one hot source, if the source(s) is(are) locally
thermalized and what are the regions accessed by these systems in the
equation-of-state phase diagram.  
From a theoretical
point of view, understanding the midvelocity emissions is linked
to the more general issue of isospin dynamics \cite{baran,ditoro}, 
which are supposed to be ruled by the symmetry-energy term.
This term is unknown at the low barion densities predicted for the
midvelocity matter when a neck develops between the interacting nuclei
during the fast separation phase \cite{xiao,rizzo}.  

Several experiments (e.g.~\cite{huda5,pia6,theri}) have shown that
the characteristics of the midvelocity emissions are very different with
respect to the emissions from the 
quasi-projectile ($QP$) [and quasi-target ($QT$)]
and standard evaporation codes can give a good reproduction of the 
$QP$ emissions, but not of the midvelocity emissions.
From an accurate energy balance \cite{mang} it was found that the $QP$
and $QT$ become more and more excited with decreasing impact parameter.
At the same time, already for peripheral impacts, the energy deposited 
in the matter emitted at midvelocity accounts for a large fraction of
the energy globally dissipated into the internal degrees of freedom;
it is possible that the resulting large excitations easily overcome
the limit of 3-4\AMeV\ for which many authors have shown the onset of
instabilities in nuclei. 
Therefore, we found appropriate
to verify the compatibility of the experimental results 
with calculations performed in the frame of the Statistical
Multifragmentation Model (SMM) \cite{botvi87,bondo,botvi01}. 

 \section{Characterization of the reaction scenario}  

The experimental data used for the comparison with the SMM 
calculations are those of the $^{93}$Nb + $^{93}$Nb collision at
38\AMeV\ \cite{pia6}.
The experimental setup, the analysis procedures
and the method for separating the midvelocity emissions from the
evaporative decays are described in detail in Refs. \cite{fiasco,pia6}.
Here we briefly remind that binary events were fully characterized by
detecting in coincidence the $QP$ and the $QT$, together with the 
associated emissions of light charged particles (LCP, $Z\leq 2$)
and intermediate mass fragments (IMF, $3\leq Z \leq 7$);
isotopic discrimination was obtained for Z=1 and charge identification 
for reaction products up to Z=7.
Different aspects of the midvelocity emissions were presented
elsewhere \cite{pia2,pia6,pia7,pia8} and their characteristics were
compared with those of the sequential decay of the two main 
reaction products, the excited $QP$ and $QT$. 

\begin{figure}[t]       
 \includegraphics[width=62mm]{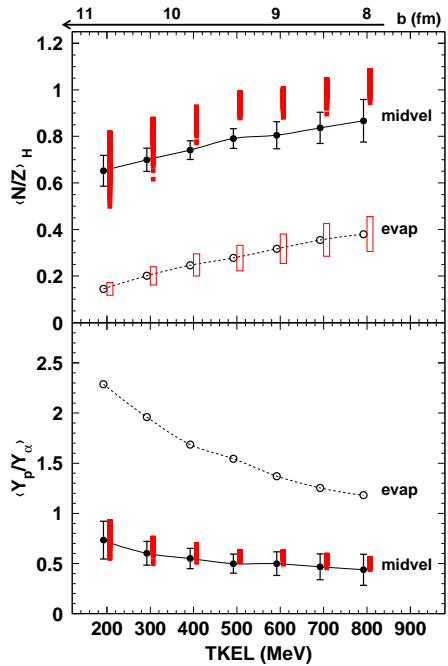}
 \caption{
   \label{f1}
  (color online)  Experimental observables for midvelocity (solid
 circles) and $QP$ (open circles) emissions as a function of TKEL:
 (a) Neutron to proton ratio for hydrogen isotopes;
 (b) Ratio of the yield of protons to $\alpha$-particles. 
 The lines through the experimental points are guides to the eye. 
 The vertical bars on the right of the experimental points 
 correspond to the range of acceptable solutions of SMM calculations 
 for the midvelocity source (solid bars) and the $QP$ (open bars), 
 as explained in the text.}
\end{figure}

All the observables related to the hot $QP$ and $QT$ are
compatible with the statistical decay of a source at normal density,
which can reach excitation energies up to about 3\AMeV\  
(T $\leq$ 6~MeV) \cite{pia6,pia7}, in agreement with the findings of 
other authors \cite{huda5,jandel,huda6,theri,bonnet}
when comparable windows of impact parameter are selected.
In particular, the LCP multiplicities and their ratios, as well as the
neutron content of hydrogen isotopes (see for instance Figs. 10-12 of
Ref. \cite{pia6}), are in good agreement with the predictions of commonly
used evaporation codes. 
(At variance with other authors, e.g. Ref.\cite{milaz}, 
we don't need to invoke a statistical multifragmentation mechanism
for the observed decay of $QP$ and $QT$).

On the contrary, the midvelocity emissions present features which make them
strongly different from the usual decay of hot rotating nuclei.  
This is clearly shown by the results of Fig.~\ref{f1}. 
Part (a) of the figure presents the average experimental isospin-ratio
of hydrogen,  $\langle N/Z \rangle_H$,
both for the midvelocity emissions (solid circles) and for the $QP$ decay
(open circles), as taken from Fig. 12 of Ref. \cite{pia6}.
Part (b) of Fig.~\ref{f1} presents the experimental ratio of the 
proton/$\alpha$-particle yields, $\langle Y_p/Y_{\alpha} \rangle$,
again for the midvelocity emissions (solid circles) and for the $QP$ decay
(open circles); this part has been derived from the multiplicities of
Fig. 5 of the same Ref. \cite{pia6}. 

The experimental results are presented as a function of a variable,
TKEL, which is used here just as an ``ordering parameter'' to classify
the events in bins of decreasing impact 
parameter \footnote{TKEL is defined as the difference between the
     initial center-of-mass energy $E_{\mathrm{in}}^{\mathrm{\;c.m.}}$
     of the collision and the c.m. kinetic energy of the outgoing $QP$
     and $QT$, assuming binary kinematics}.
This use of the variable TKEL has been extensively discussed in a
previous paper~\cite{pia6}, where it was shown that 
- even at Fermi energies - TKEL is strictly correlated with the relative
velocity between $QP$ and $QT$ and, as such, it is an indicator of the
violence of the collision (and hence of the impact parameter);  
this interpretation has been recently adopted also in a theoretical
description of the reaction dynamics~\cite{rizzo}.
The estimated correspondence of TKEL with impact parameter 
for the data of this paper is displayed by the arrow at the top margin
of Fig.~\ref{f1}.  
It shows that the presented data refer to semi-peripheral collisions in
the impact-parameter range between 8 and 11 fm. 
More central events have not been used because, with decreasing relative
velocity, the separation between midvelocity emissions and $QP$ (or $QT$)
evaporation becomes more and more uncertain and hence the interpretation
of the results would become less and less reliable.

From the data of Fig. \ref{f1} it is apparent that the midvelocity
emissions present some substantially different characteristics with
respect to the usual evaporative processes. 
While the $\langle N/Z \rangle_H$ ratio for particles
emitted by the $QP$ (open circles) is rather low and it was found
compatible with the results of \textsc{Gemini} calculations for the
decay of a $^{93}$Nb nucleus at normal density with the appropriate
excitation energy \cite{pia6}, the same ratio for the midvelocity
emissions (solid circles) is much higher and cannot be reproduced by
usual statistical model calculations. 
Even more striking is the fact that the relative abundances of the 
emitted particles are different, displaying some remarkable inversions
\cite{pia6}; 
in particular Fig. \ref{f1}(b) shows that the relative abundances of
protons and $\alpha$ particles are reversed when going from the
evaporative to the midvelocity emissions.

The exact nature of the midvelocity emissions is yet not fully
ascertained. 
On the one hand, one observes a component which is
evidently Coulomb-related \cite{huda6,pia7} to the main reaction
products: this suggests that they are driven by the reaction dynamics
towards rather elongated shapes, which quickly decay before complete
decoupling from the formation stage \cite{pia2,huda6}.   
On the other hand, there are the very 'central' emissions, located at 
small velocities in the c.m. system: they may be attributed to the
disassembly of the overlap region (possibly characterized by high
excitations and small densities), or to the snap-off of distended
neck-like structures formed just in between the two separating main
products~\cite{huda5}.   
In any case, our previous results~\cite{mang,pia7} suggest that the
energy density in this region may reach values well inside the region
in which nuclear multifragmentation signals have been detected. 
Dynamical codes predict the formation of
elongated structures in fast dissipative collisions in the Fermi energy
domain~\cite{ditor1,baran,huda6,rizzo}.
As to the timescale of the formation/emission processes, dynamical
calculations \cite{ditor1,baran,rizzo} for semiperipheral reactions
suggest that it should be short (about 100-200~fm/c), but still
long enough to allow the system to reach at least a partial
equilibration in some degrees-of-freedom at a freeze-out stage. 
The final conclusion about statistical 
equilibrium can be obtained only with a comprehensive comparison 
with experiments.

Thus, an hypothetical ``source'' of the midvelocity emissions 
-- supposing that it exists -- may
have a more complex configuration than that considered in
the frame of a statistical multifragmentation model.
Therefore, at variance with what one expects in central collisions,
in peripheral collisions we are aware that SMM calculations 
may not reproduce all details of the midvelocity 
emissions, especially concerning the kinematical characteristics of 
fragments, where the role of dynamics is still quite important.
Nevertheless, the chemical equilibrium, responsible for the isospin
composition of the fragments produced in the source, may be established.
If the model is able to capture -- at least partly -- some
important characteristics of the process (which might be, e.g., the
tendency of some portion of nuclear matter to rapidly decay after having
reached a partial thermal and chemical equilibrium, or possibly some
kind of dilution being subjected to mechanical strains), 
one may expect that the model displays reduced discrepancies with the
experimental data than the usual low-energy statistical models.
With this in mind, is seems appropriate to test the compatibility of
statistical multifragmentation
calculations with the measured properties of the midvelocity emissions.

   \section{The parameters of the SMM calculations}

The Statistical Multifragmentation Model (SMM) is a well developed code
which describes the nuclear disassembly in various 
regimes~\cite{botvi87,bondo,botvi01}. 
Present SMM calculations include also the decay of the hot fragments
produced at the freeze-out, via evaporation or Fermi break-up
\cite{botvi87}. 
Therefore we can directly compare the ``final state'' quantities
computed by the code with the experimental ones, which are by
definition secondary, i.e. after sequential particle decay. 
For the comparison with the data, we consider here the two experimental 
variables, $\langle N/Z \rangle_H$ and $\langle Y_p/Y_{\alpha} \rangle$,
already presented in the previous Section:
being relative quantities, they are less sensitive to systematic
uncertainties, both from the experimental and theoretical point of view. 
Other authors~\cite{xu} used other experimental variables
for a comparison with SMM.

In peripheral collisions the input values for the SMM calculations
are by far less obvious than in the usual applications of the model, 
which mostly refer to a single source produced in central collisions 
\cite{geraci,bonnet}. 
There are no reliable hints concerning the appropriate values for the
source size $A_s$, its charge $Z_s$ and its excitation energy per
nucleon $\epsilon_s$, as well as their dependence on impact parameter. 
In fact, besides the uncertainties due to the usual lack of information
about the emitted free neutrons, in peripheral collisions the process
may be more complicated than the simple emission from a source: 
for example, part of the ejected matter could be reabsorbed by the $QP$
and $QT$ which, while flying apart, are still rather close to each other
during the midvelocity emission phase~\cite{pia7}. 

Given these uncertainties on the source parameters, we decided to run
many SMM calculations with a wide grid of input values for $A_s$,
$Z_s$ and $\epsilon _s$. 
A lower limit for the charge of the source, $Z_s$, was given by the 
experimental total charge of the midvelocity emissions, 
$\langle Z_{midv} \rangle$, 
which smoothly increases with TKEL, up to about 12--15 for the highest 
TKEL values considered in this paper.
In other words, the lower limit for $Z_s$ corresponds to the extreme
assumption that the source completely disassembles into the measured
total charge $\langle Z_{midv} \rangle$.  
As an upper limit, one might have taken the overlap matter in a
participant-spectator picture 
(with or without taking into account the diffuseness);
however, in order not to bias the results with a too restricted choice,
we preferred to use in all cases a value of 50 charge units, which is
rather large (it corresponds to more than half of the total charge of
our system and we recall that we are dealing with peripheral collisions), 
checking afterwards the sensitivity of the results on this parameter.

The second input parameter is the mass of the source, $A_s$.
Multifragmentation processes are much more sensitive to
the neutron content of the source, $(N/Z)_s \equiv (A_s-Z_s)/Z_s$,
than to the source size itself.
In fact the isospin dynamics are supposed to play a relevant role in the
fragment production, either via the density dependence of the symmetry
energy~\cite{ditor1,baran} or via polarization effects of the nuclear
matter due to the Coulomb field \cite{botvi99,jandel}.
Our symmetric system has an initial N/Z value of 1.27, but we have no
other clue on the isospin evolution during the interaction and hence on
the N/Z values appropriate for the various portions of the system.
If no free midvelocity neutrons were emitted at all, from the
experimental data one would obtain for the midvelocity emissions
a low (maybe unrealistic) value of N/Z=1.02-1.06 at $TKEL\ge$400MeV.
Therefore, again in order not to bias the results, for each $Z_s$ we
allowed a large variation of the source size $A_s$, so that the resulting 
ratio $(N/Z)_s$ was comprised in the wide range between 0.9 and 2.0. 

Concerning the excitation of the source, $\epsilon_s$, the experimental 
value obtained via calorimetry -- assuming a complete disassembly of the
midvelocity source into midvelocity emissions, plus some neutrons 
(Fig.3 in Ref.~\cite{mang} or Fig.9 in Ref.~\cite{pia6}) -- points to
high excitations for the overlapping matter, well above the values
deduced for the $QP$ and $QT$. 
In the calculations we let $\epsilon_s$ span the range from 4 up to 
9~MeV, which approximately corresponds to the vaporization limit.
We neglect any radial-energy contribution to the excitation, as it has
been shown that the compression-expansion cycle is weak in peripheral
reactions \cite{bonnet}.
Moreover SMM results on fragment yields are
insensitive to radial flow \cite{xu}.

Another parameter of SMM calculations is the freeze-out density of the
source.  
Values of $\rho$ ranging from  $\rho_0/6$ \cite{xu} to
$\rho_0/3$~\cite{dagoauau,milaz,geraci,shet} have been commonly used
in SMM calculations (with $\rho_0$ normal nuclear density).
The possibility that dilute systems are formed also in semiperipheral
collisions has been suggested by BNV calculations~\cite{baran} which
show that, for a system of similar size ($^{124}$Sn + $^{58}$Ni at
35\AMeV\ (b=6~fm)), very low densities can be attained in the neck
region within 160-250~fm/c, especially when a soft equation-of-state is 
assumed.
We run SMM with the value $\rho=\rho_0 /6$ and we verified that the
results do not appreciably change when increasing the density 
up to $\rho=\rho_0/3$.

A last word on the source shape: the inclusion of deformed shapes for
the source modelization is in principle relevant. 
As already said, elongated transient nuclear systems are likely to be
formed in semiperipheral collisions; recently, exotic shapes have been
suggested even for central event sources, thus leading to
multifragment calculations for non-spherical configurations~\cite{lefev}. 
However we think that our data concerning isotopic yields of lightest
fragments 
are insufficient to investigate with SMM also this degree-of-freedom,
which therefore has been neglected here (only spherical geometry is
assumed).

   \section{Comparison with the data and discussion}  

Among the many triples ($A_s, Z_s, \epsilon_s$) defining the sources on
the grid for the SMM calculations, we retained only those for which both
calculated values of $\langle N/Z \rangle_H$ and 
$\langle Y_p/Y_{\alpha} \rangle$ differed from the
corresponding experimental values by less than 25\%.
Out of over 5000 triples used for the input grid, only about 200 were
found to pass the above mentioned criterion of goodness at each TKEL.
The values of $\langle N/Z \rangle_H$ and 
$\langle Y_p/Y_{\alpha} \rangle$ for the so defined ``good'' 
SMM sources are shown, for each TKEL, by squares in Fig.~\ref{f1}.
Since all the ``good'' sources tend to give values of these two
parameters quite close to each other, the squares tend to bunch up into
the vertical solid bars which are visible in Fig.~\ref{f1}, slightly
shifted -for the sake of clarity- to the right of each experimental point. 

It is worth noting that many slightly different sources, all with rather
reasonable parameters, are able to simultaneously reproduce both
experimental correlations of Fig. \ref{f1}, including their dependence
on impact parameter; 
this is a result which was {\bf not} obvious {\it a priori}.  

As it was already shown in Ref. \cite{pia6}, the midvelocity emissions
are appreciably richer in deuterons and tritons (with respect to the 
protons) than the evaporative emissions,
a feature which is reproduced by the SMM calculations.
The results of the SMM calculations tend to be even more n-rich than the
experimental data, but the discrepancy is not very large
\footnote{We note that the midvelocity yields, obtained
       in \cite{pia6} by subtracting the evaporative component from
       the total, actually refer to all not-equilibrated emissions,
       including those which are anisotropically distributed on the
       Coulomb ridges of the $QP$ and $QT$ \cite{huda4,pia7} and
       might be less exotic in their neutron content.}.
Also the inversion in the relative abundances of protons and
$\alpha$-particles between midvelocity and evaporative data, which is  
displayed by the ratio $\langle Y_p/Y_{\alpha} \rangle$ in
Fig. \ref{f1}(a), is well reproduced by the same ``good'' SMM
calculations. 

This comparison shows that the examined features of midvelocity
emissions bear a closer resemblance with the expectations of a
multifragmentation model than with those of a sequential decay.
To give more support to this last statement, we note that, in
its recent versions, the SMM code has been upgraded to treat also the
'standard' evaporation of hot nuclei at moderate excitations and normal
density.   
Therefore we have run the SMM code also in this mode (with input
parameters similar to those used for previous GEMINI calculations, see
Fig. 12 of Ref. \cite{pia6}) and also this comparison with the
experimental data for the $QP$ decay is presented by the open bars 
to the right of the open circles in Fig.~\ref{f1}(a).   
It is worth noting that the large difference in $\langle N/Z \rangle_H$
between evaporative and midvelocity emissions is well reproduced by SMM.
We did not try to calculate
$\langle Y_p/Y_{\alpha} \rangle$ for the evaporation of the $QP$ 
within SMM, since this ratio strongly depends on the angular momentum 
of the emitting source \cite{casini1,casini2}. 
This would require the determination of angular momentum as a function
of the impact parameter for the $QP$ source and it would divert from the
main subject of the paper, which concerns the midvelocity source.

It may be worth examining in more detail the parameters of the ``good''
sources which lead to the observed agreement with the experimental
points for the midvelocity emissions (circles in Fig. \ref{f1}).
For the source size $A_s$, rather flat distributions are obtained, thus
indicating the insensitivity of the presently considered isospin
quantities to this SMM parameter;
a similar insensitivity of other experimental quantities to the source
size was already noted in Refs. \cite{xu,huda5}.  
Concerning the energy density, $\epsilon_s$, in all cases the best
agreement is obtained for rather large excitation energies, well inside
the multifragmentation region:   
most good triples have $\epsilon_s$ in the range 7--9 MeV, 
with a very weak tendency to increase with increasing TKEL.

\begin{figure}[t]       
 \includegraphics[width=70mm]{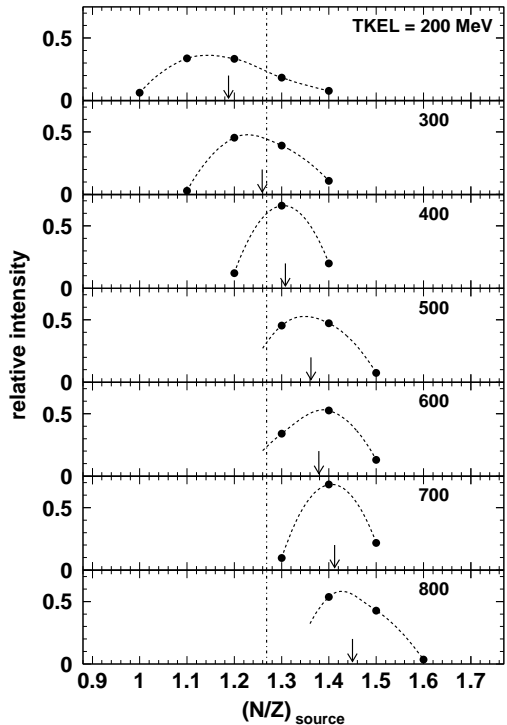}
 \caption{
   \label{f2}
 $(N/Z)_s$ distributions for the midvelocity multifragmenting sources of
 the SMM calculations which produce the results shown in Fig. \ref{f1}.
 From top to bottom, the TKEL windows go from very peripheral to 
 mid-peripheral events.
 Each distribution is normalized to unit area and the arrow shows 
 its average value; the vertical line is the $N/Z$ of the system.}  
\end{figure}

Coming to the neutron content, the $(N/Z)_s$ of the ``good'' sources is
presented as a function of $TKEL$ in Fig.~\ref{f2}, where all
distributions have been normalized to unity. 
It is interesting to note that the average value of $(N/Z)_s$ (indicated
by the arrows in Fig. \ref{f2}) displays a definite increasing trend
with increasing TKEL.  
Only in the most peripheral bin it remains below that
of the colliding system (1.268, indicated in figure by the vertical
line), while for the other less peripheral bins it rises and reaches  
values around and above 1.4.
If the source size increases with decreasing impact parameter, this might
naturally tend to privilege higher $N/Z$ values. 
To exclude this trivial explanation, we checked that no appreciable
correlation exists between $(N/Z)_s$ and $A_s$;  
in addition, limiting the accepted source sizes to 40 amu does not change
the trend observed in Fig. \ref{f2}.
One could have attributed this trend to the required agreement with 
$\langle N/Z \rangle_H$, an experimental observable which strongly
depends on deuterons: because of their more diffuse wave function,
deuteron production could be enhanced in too diluted configurations
\cite{xu}.
However, it was verified that increasing the freeze-out density to
$\rho=\rho_0/3$ produces negligible effects on the behaviour of $(N/Z)_s$.

The question of a possible neutron enrichment is widely discussed
in the literature.
Theoretically, Landau-Vlasov calculations with different chemical
potentials for neutrons and protons predict -for the neck matter-
a neutron enrichment with respect to the initial value when an asy-stiff
equation-of-state is used \cite{ditor1,baran,rizzo}.
For example, in a recent work \cite{rizzo}, several effects contributing
to isospin dynamics in semiperipheral reactions were studied.   
The predicted n-enrichment of the neck-matter occurs, even for symmetric
systems, via the socalled ``isospin migration'' process, which sets in
because of the density gradient between the $QP$ (and $QT$) and the more
diluted neck matter.
The observed rising trend of $(N/Z)_s$ with increasing 
centrality is in agreement with other experimental indications
\cite{theri}. 
The rather peripheral impact parameters addressed in this work
($b \approx$ 8--11 fm) do not allow to check whether the N/Z of the 
midvelocity matter decreases for substantially more central
collisions, as it is predicted by some dynamical calculations
(see, e.g., \cite{baran02}).

Experimental results are not easy to compare with each other and they 
usually give indications based on the isotopic analysis of only some 
species (IMFs or LCPs) emitted by the multifragmentation of the source.
These partial results are usually interpreted as indications that,
already in peripheral collisions, the midvelocity source is 
neutron enriched with respect to the the $QP$ source 
(see e.g. \cite{milaz,xu,huda5,pla}).
Complete measurements of all kinds of products emitted at midvelocity 
are rare. 
Combining the data of two experiments, one about the free neutrons and
the other about the charged products, it was concluded \cite{sobo}
that globally all the material found in the midvelocity region is likely
to have the same N/Z ratio as the bulk matter.
However, more recently, a simultaneous isotopic analysis of all emitted
products (IMFs, LCPs and neutrons) has been performed \cite{theri},
leading to the opposite conclusion:
although the errors are very large, it gives support to the idea that
the midvelocity matter is more n-rich than the initial system and it
tends to become even more n-rich with decreasing impact parameter.

In the present case, the results of Fig. \ref{f2} are just a
model-dependent indication that some of the features observed at
midvelocity point to a possible neutron enrichment of the neck region,
once a statistical multifragmentation mechanism is assumed 
to be responsible for this kind of emissions.   
Whether this possible neutron enrichment is mainly produced by density
dependent effect\cite{rizzo,ditoro,baran}, or by a lowering of the 
symmetry energy in the dilute hot multifragmenting systems\cite{souli},
or by some other exotic mechanism, is still an open and widely debated
question, which deserves further investigation.

   \section{Conclusions}
  
In semiperipheral collisions ($b \ge $8~fm) at Fermi energies it was
found that the experimentally observed intense emission of reaction
products from the midvelocity region displays some characteristics which
are quite different from those of the usual evaporation from hot nuclei.
The excitation energy per nucleon in the overlap region has been
estimated to be well above the commonly accepted limit for 
multifragmentation.
Therefore, we have investigated in how far the behavior of the above
mentioned experimental quantities may be reproduced by calculations in
the frame of the Statistical Multifragmentation Model (SMM).
The main outcome of this investigation is that it is indeed possible 
to find source parameters such that the final
multifragmentation products reproduce reasonably well the observed
peculiar features of midvelocity emissions. 
In particular they reproduce the neutron enrichment of the emitted
hydrogen isotopes, as well as the inversion of relative abundances of
protons and $\alpha$-particles with respect to an evaporative process.
These SMM sources, while showing no preference for a particular size
$A_s$, are characterized by rather high 
excitation energies (7--8 MeV per nucleon)
and by a tendency to become more neutron-rich than the
colliding system with decreasing impact parameter.


\vspace{1mm}

  \begin{acknowledgments}

 One of the authors (A.B.) acknowledges financial support from 
the FAI fund of INFN.
 \end{acknowledgments}
%
%
%

  \bibliography{smm}
%

\end{document}